\documentclass{ismdproc}

\usepackage{epsfig}
\usepackage{cite}

\begin{document}
\title{Recent developments in NLO QCD calculations: the particular
  case of $pp \rightarrow t\bar tjj$\footnote{WUB/10-38}}



\author{{\slshape Ma\l{}gorzata Worek\footnote{Presented at the XL 
International Symposium on Multiparticle Dynamics, ISMD 2010, 
21-25 September 2010, Antwerp, Belgium.}} \\[1ex]
Fachbereich C Physik, Bergische Universit\"at Wuppertal, D-42097 
Wuppertal,Germany}

\contribID{xy}  
\confID{yz}
\acronym{ISMD2010}
\doi            

\maketitle

\begin{abstract}

A very brief status of next-to-leading order QCD calculations is given.
As an example the  next-to-leading order QCD calculations to the
$pp\to t\bar{t}jj$ 
processes at the CERN Large Hardon Collider are presented.  Results
for integrated and differential cross sections are shown. They
have been obtained in the framework of the \textsc{Helac-Nlo} system.

\end{abstract}

\section{Introduction}

At the CERN Large Hardon Collider we hope to uncover the mechanism of 
electroweak symmetry
breaking  and to find signals of physics beyond the Standard Model.
Signal events have to be dug out from a bulk of background  events,
which are due to the Standard Model processes, mostly QCD processes
accompanied by additional electroweak bosons. These constitute of
final states  with a high number of jets or identified particles. LHC
data can only be meaningfully analyzed if a plethora of Standard Model
background processes  are theoretically under control.  If one is
content with a leading order description of the multijets final states
it is possible to go to quite high orders, say 8-10 particles in the final
states, which have to be well separated to avoid phase space regions
where divergences become troublesome. Quite a number of tools enable
one to do this: 
\textsc{Alpgen}\footnote{{\tt http://mlm.home.cern.ch/mlm/alpgen/}} 
\cite{Mangano:2002ea},
\textsc{Amegic++/Sherpa}\footnote{\tt http://www.sherpa-mc.de/} 
\cite{Krauss:2001iv,Gleisberg:2008ta},
\textsc{Comix/Sherpa}\footnote{\tt http://www.freacafe.de/comix/} 
\cite{Gleisberg:2008fv}, 
\textsc{CompHEP}\footnote{\tt http://comphep.sinp.msu.ru/}
\cite{Pukhov:1999gg}, \textsc{Helac-Phegas}
\footnote{\tt http://helac-phegas.web.cern.ch/helac-phegas/}
\cite{Kanaki:2000ey,Papadopoulos:2000tt,Cafarella:2007pc},
\textsc{MadGraph/MadEvent}\footnote{\tt http://madgraph.hep.uiuc.edu/} 
\cite{Maltoni:2002qb,Alwall:2007st} and
\textsc{O'mega/Whizard}\footnote{\tt http://projects.hepforge.org/whizard/} 
\cite{Kilian:2007gr}.  Those tools, which are based on  Feynman
diagrams  suffer from inefficiency at large n, where n is a number of
particles, because 
the number of diagrams to be evaluated grows rapidly as n
increases. Some of the tools above use methods designed to be
particularly efficient at 
high multiplicities. Namely, they build up amplitudes for complex
processes using off-shell recursive methods.  Nevertheless, all these
tools are completely self contained and provide amplitudes and
integrators on their own.  Although multijet observables can rather
easily be modeled at leading order,  this description suffers several
drawbacks. Leading order calculations depend strongly on the
renormalisation scale and can therefore give only an order of
magnitude estimate on absolute rates. Besides normalization, sometimes
also  shapes of distributions are first known at higher  orders.
Secondly, for many scale processes like {\it e.g.}  $t\bar{t} H$,
$t\bar{t} + nj$, $nj$,  $V +nj$,  $VV + nj$,  where $V$ stands for
$W^{\pm}$ and/or $Z$, a proper scale choice is problematic. For
some observables dynamical scales seem to work better, for others
fixed scales are applied. How do we know which scale to choose ?
Moreover, at leading order a jet is modeled by a single parton, which
is a very crude approximation. The situation can significantly be
improved by including higher order corrections in perturbation theory.
Next-to-leading order programs can be divided into three categories.
Libraries with a specific list of  processes at hadron-hadron
colliders like {\it e.g.}  \textsc{Mcfm}\footnote{\tt http://mcfm.fnal.gov/} 
for processes
with heavy quarks and/or heavy  electroweak bosons, 
\textsc{NloJets++}
\footnote{\tt http://www.desy.de/$^{\sim}$znagy/Site/NLOJet++.html}
\cite{Nagy:2003tz} for jet production  and 
\textsc{Vbfnlo}
\footnote{\tt http://www-itp.particle.uni-karlsruhe.de/$^{\sim}$vbfnloweb/}
\cite{Arnold:2008rz} for vector-boson fusion processes.  Automatic
tools based on Passarino-Veltman \cite{Passarino:1978jh} 
reduction of one-loop  amplitudes for general $2\to n$ processes   
like {\it e.g.} 
\textsc{FeynArts/FormCalc/LoopTools}\footnote{\tt http://www.feynarts.de/} 
\cite{Hahn:1998yk,Hahn:2000kx} and 
\textsc{Golem}\footnote{\tt http://lapth.in2p3.fr/Golem/golem95.html}
\cite{Binoth:2008uq}. And finally, automatic tools based on OPP
reduction   and other unitarity-based methods, which
aim at multiparticle  processes at hadron colliders like
\textsc{BlackHat/Sherpa},  \textsc{Rocket/Mcfm} and \textsc{Helac-Nlo}.
Thanks to all these methods and developments several $2\to 4$
processes  have recently been calculated at next-to-leading order
 QCD, including  $pp
\to t\bar{t}b\bar{b}$ \cite{Bredenstein:2009aj,
  Bevilacqua:2009zn,Bredenstein:2010rs},  $pp \to W^{\pm} + 3j$
\cite{Ellis:2009zw,Berger:2009zg,
  KeithEllis:2009bu,Berger:2009ep,Melnikov:2009wh},  $pp(q\bar{q}) \to
b\bar{b}b\bar{b}$ \cite{Binoth:2009rv}, $pp \to t\bar{t}jj$
\cite{Bevilacqua:2010ve},  $pp \to Z\gamma^{*} + 3j$
\cite{Berger:2010vm},  $pp \to W^\pm W^\pm jj$ 
 production via  weak-boson fusion
\cite{Jager:2009xx},   QCD-mediated $pp \to W^+W^+ jj$
process \cite{Melia:2010bm},   and finally $pp \to W^{+}W^{-}
b\bar{b}$ \cite{Denner:2010jp,Bevilacqua:2010qb}.  
Additionally, the first $2 \to 5$ process $pp \to W^{\pm} +
4j$ \cite{Berger:2010zx} has recently been calculated in the
leading-color approximation.

In this contribution, a brief report on the \textsc{Helac-Nlo}  approach 
and the $pp \to t\bar{t}jj$ computation is given.

\section{Details of the next-to-leading order calculation}

The next-to-leading order   results are obtained in the framework of
\textsc{Helac-Nlo} based on the \textsc{Helac-Phegas} leading-order
event generator for all parton level, which has, on its own, already
been extensively used and tested in phenomenological studies see {\it
  e.g.}  \cite{Gleisberg:2003bi,Papadopoulos:2005ky,Alwall:2007fs,
  Englert:2008tn,Actis:2010gg}.  The integration over the fractions
$x_1$ and $x_2$ of the initial partons is optimized with the help of
\textsc{Parni}\footnote{\tt http://helac-phegas.web.cern.ch/helac-phegas/parni.html} 
\cite{vanHameren:2007pt}. The phase space integration
is executed with the help of  
\textsc{Kaleu}\footnote{\tt http://helac-phegas.web.cern.ch/helac-phegas/kaleu.html} 
\cite{vanHameren:2010gg}
and cross checked with \textsc{Phegas} \cite{Papadopoulos:2000tt},
both general purpose multi-channel phase space generators.   The
next-to-leading order system consists of:
\begin{enumerate}
\item \textsc{CutTools}\footnote{\tt http://www.ugr.es/$^{\sim}$pittau/CutTools/} 
\cite{Ossola:2007ax}, for the OPP reduction
  of tensor integrals with a given numerator to a basis of scalar
  functions and for the rational parts
  \cite{Ossola:2006us,Ossola:2008xq,Draggiotis:2009yb}; 
\item \textsc{Helac-1Loop} \cite{vanHameren:2009dr} for the evaluation
  of one loop amplitude, more specifically for the evaluation of  the
  numerator functions for given loop momentum (fixed by
  \textsc{CutTools}); 
\item  \textsc{OneLOop}\footnote{\tt http://helac-phegas.web.cern.ch/helac-phegas/OneLOop.html} 
\cite{vanHameren:2009dr,vanHameren:2010cp},  
  a library of scalar functions, which provides the actual numerical 
  values of  the integrals. 
\item \textsc{Helac-Dipoles}\footnote{\tt http://helac-phegas.web.cern.ch/helac-phegas/helac-dipoles.html} 
\cite{Czakon:2009ss}, automatic
  implementation  of  Catani-Seymour dipole subtraction
  \cite{Catani:2002hc}, for the  calculation of the real emission
  part.
\end{enumerate}
Let us emphasize that all parts are calculated fully numerically   in
a completely automatic manner. 

\section{Numerical Results}

\begin{figure}
\begin{center}
\includegraphics[width=0.49\textwidth]{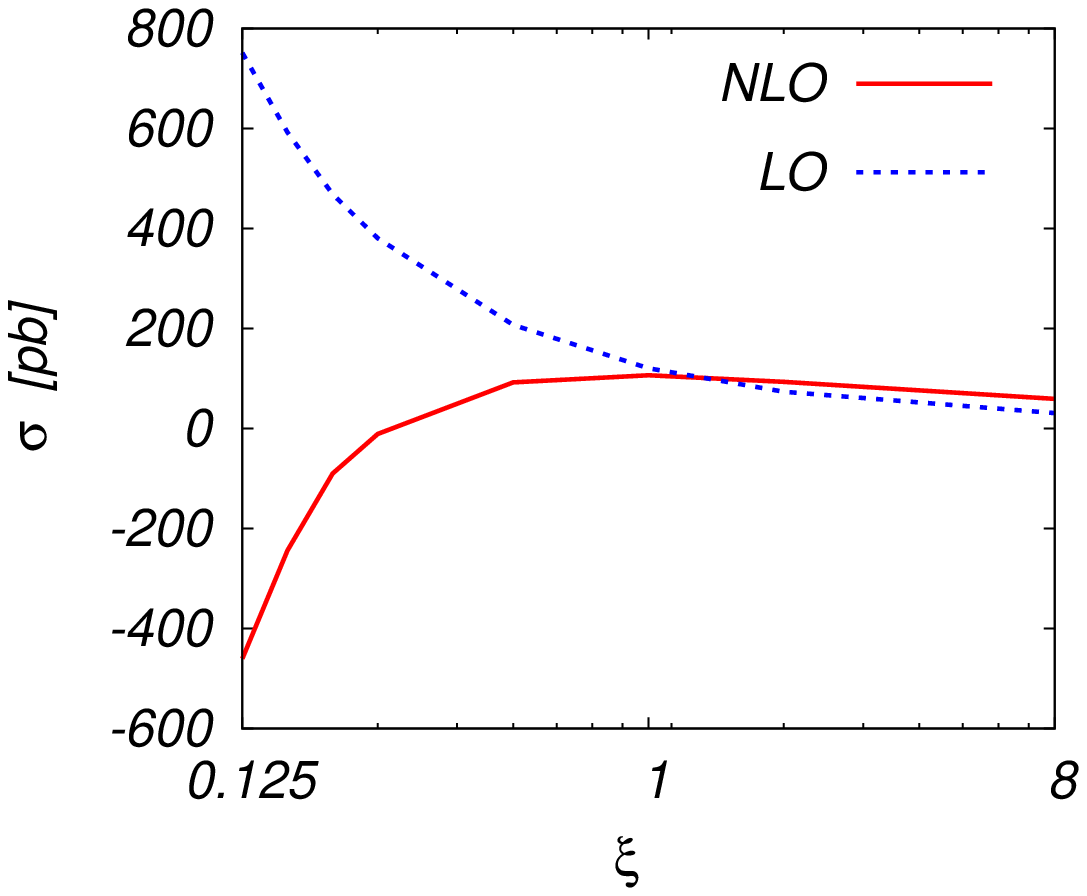} 
\includegraphics[width=0.49\textwidth]{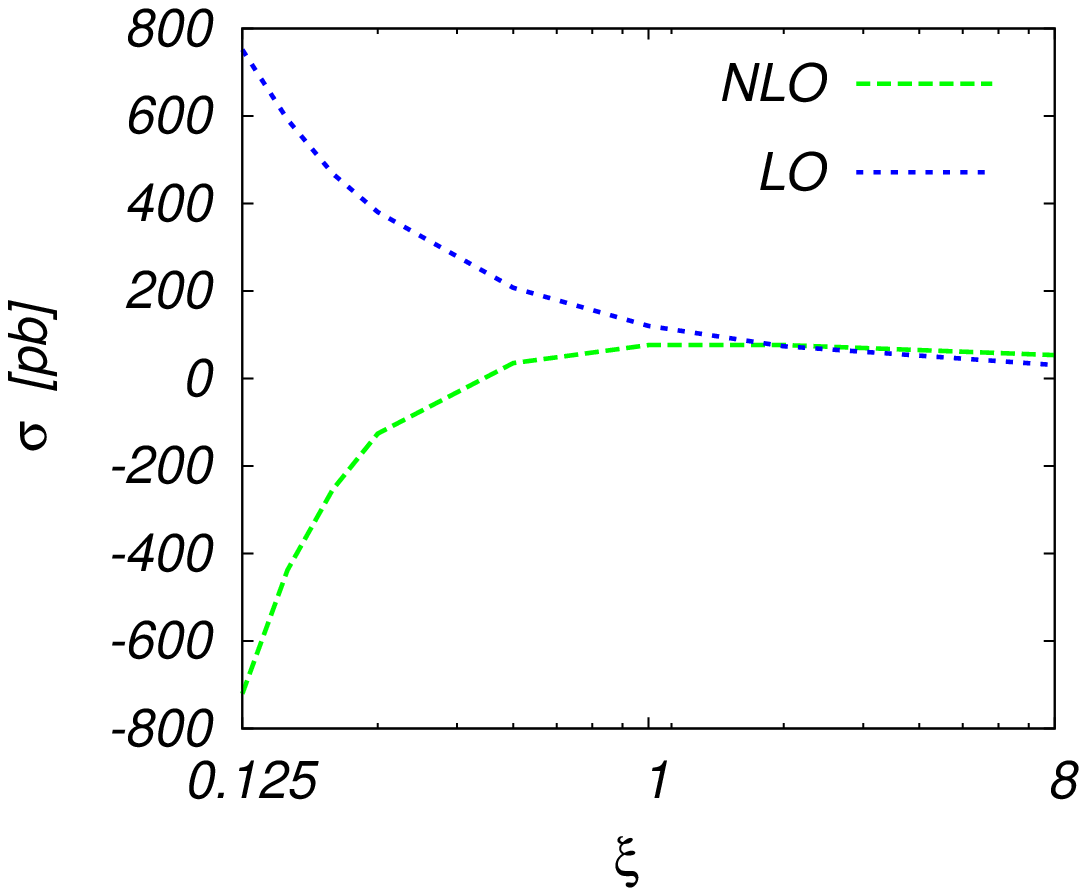} 
\caption{\it Scale dependence of the total cross section for
  $pp\rightarrow t\bar{t} jj + X$ at the LHC   with $\mu_R=\mu_F=\xi
  \cdot m_t$.  Left panel: The blue dotted curve corresponds to  the
  leading order whereas the red solid to the next-to-leading order
  one. Right panel: The blue dotted curve corresponds to  the leading
  order whereas the green dashed to the next-to-leading  result with a
  jet veto of 50 GeV.
\label{fig:scales}}
\end{center}
\end{figure}
\begin{figure}
\begin{center}
\includegraphics[width=0.49\textwidth]{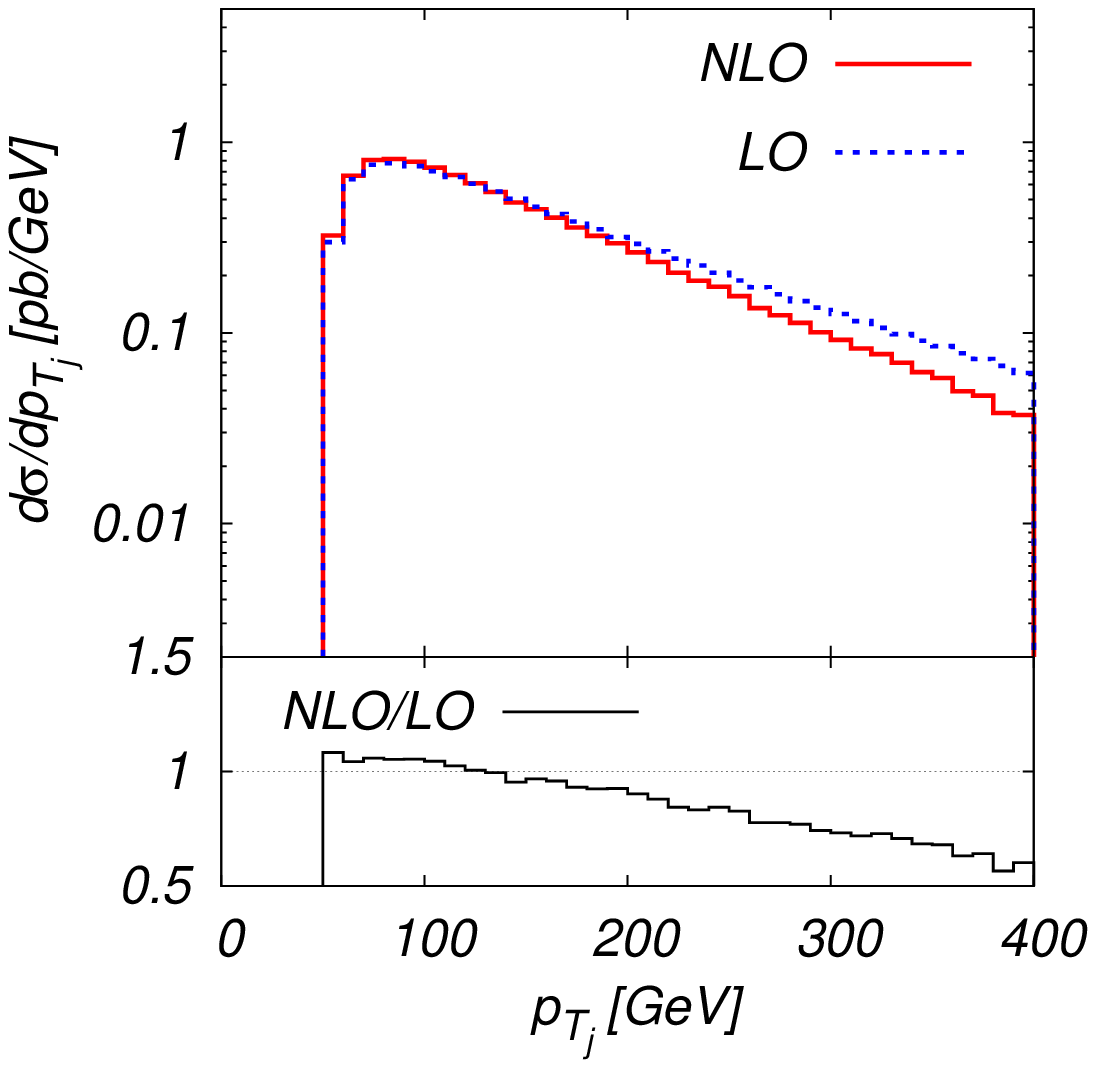}
\includegraphics[width=0.49\textwidth]{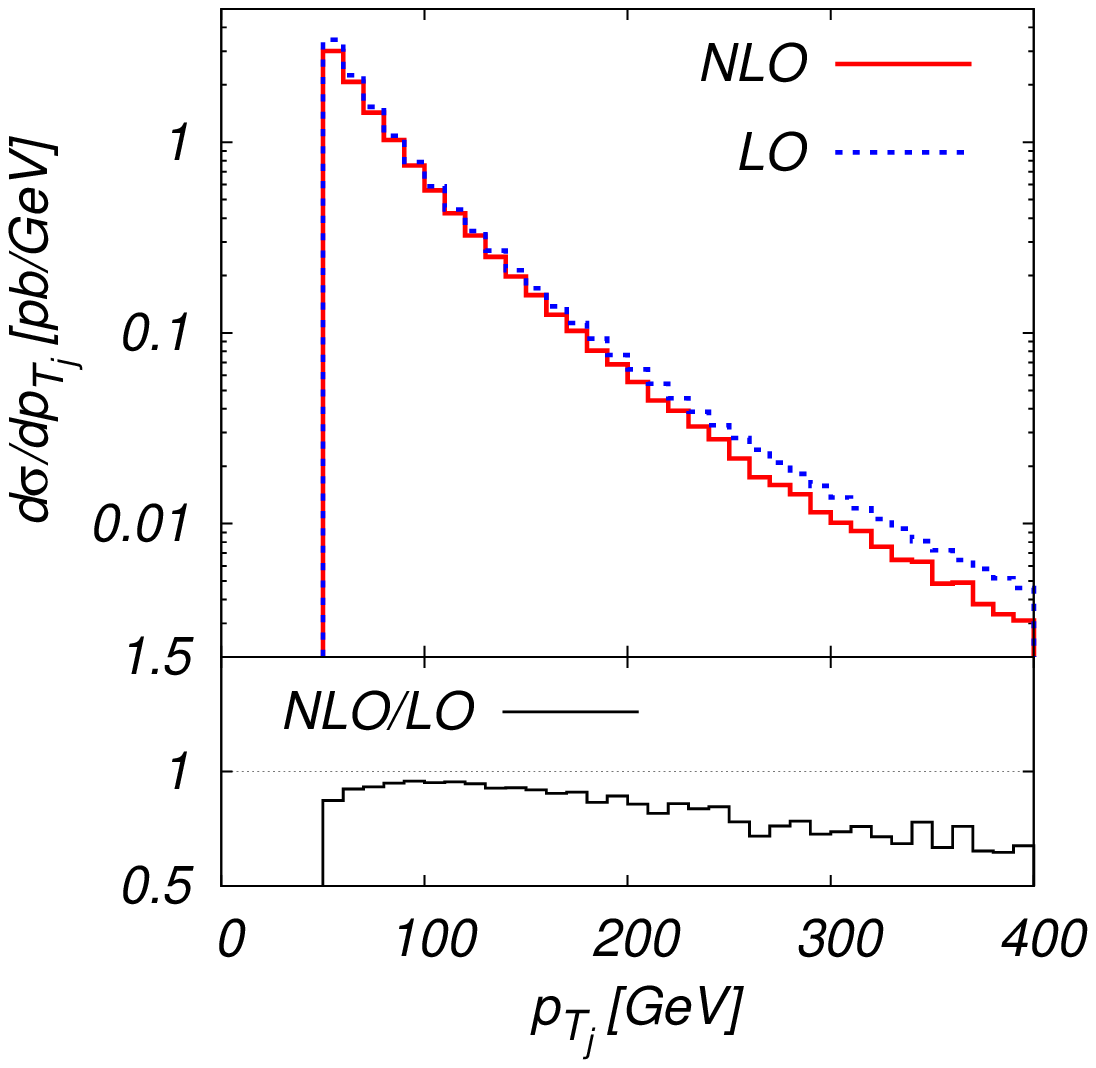} 
\caption{\it
Distribution in the transverse momentum $p_{T_{j}}$  of the   $1^{st}$
hardest jet (left panel) and the $2^{nd}$ hardest jet (right panel)  for
$pp\rightarrow t\bar{t} jj +X$ at the LHC.  The blue dotted curve
corresponds to  the leading order  whereas the red solid to the 
next-to-leading order  one.
\label{fig:ttjj}}
\end{center}
\end{figure}

We consider proton-proton collisions at the LHC with a center of mass energy
of $\sqrt{s}=14$ TeV.  The mass of the top quark is set  
to be $m_t=172.6$ GeV. We leave it on-shell with unrestricted 
kinematics. The jets are defined by at most  two partons using the $k_T$ 
algorithm \cite{Catani:1992zp,Catani:1993hr}, with a separation 
$\Delta R=0.8$, where 
\begin{equation}
\Delta R_{ij}=\sqrt{(y_i-y_j)^2+(\phi_i-\phi_j)^2}, ~~~~~~
y_i=1/2\ln\left[ (E_i-p_{i,z})/(E_i+p_{i,z})\right]
\end{equation} 
being the rapidity and $\phi_i$ the
azimuthal angle of parton $i$. Moreover, the recombination is only performed
if both partons satisfy $|y_i|<5$ (approximate detector bounds). We further
assume that the jets are
separated by  $\Delta R=1.0$ and have $|y_{\rm{jet}}| < 4.5$.  
Their transverse momentum is required to be larger than $50$ GeV.  
We consistently use the CTEQ6 set of parton distribution
functions \cite{Pumplin:2002vw,Stump:2003yu}, {\it i.e.} 
 we take CTEQ6L1 PDFs with a 1-loop running $\alpha_s$ in leading order
and CTEQ6M PDFs with a 2-loop running $\alpha_s$ at next-to-leading order.

We begin our presentation of the final results of our analysis with a
discussion of the total cross section. For the central value of the
scale, $\mu_R=\mu_F=m_t$, we have obtained:
\begin{equation}
\sigma_{\rm{LO}} = (120.17 \pm 0.08) ~{\rm pb}\, , ~~~~~~ \sigma_{\rm{NLO}} = 
(106.95  \pm 0.17) ~{\rm pb}\, .
\end{equation}
From the above result one can read a $K$ factor $K=0.89$ 
which corresponds to the negative  corrections of the order of $11\%$. 

The scale dependence of the corrections is illustrated in  Figure
\ref{fig:scales}.  We observe a dramatic reduction of the  scale
uncertainty while going from leading order to next-to-leading order.
Varying the scale up and down by a factor 2 changes the cross section
by  $+72\%$ and $-39\%$ in the leading order case, while in the
next-to-leading order case we have obtained a variation of $-13\%$ and
$-12\%$.  The third jet, which stems from real radiation, has  not
been  restricted in this  case. 

Therefore, in the next step, we study the impact of a jet veto on  the
third jet, which is simply an upper bound on the allowed  transverse
momentum, $p_T$.  The total cross section with a jet veto  of 50 GeV
is
\begin{equation}
\sigma_{\rm{NLO}}(p_{T,X} < 50 ~ \rm{GeV}) = (76.58 \pm 0.17) ~\rm{pb}
\; ,
\end{equation}
which corresponds to $K=0.64$ and negative  corrections of the order of
$36\%$. In this case a scale variation of $-54\%$ and $-0.3\%$ has
been  reached, see Figure \ref{fig:scales} for graphical
representation.

While the size of the corrections to the total cross section is
certainly interesting, it is crucial to study the corrections to the
distributions.  In Figure \ref{fig:ttjj} the transverse momentum
distributions of the hardest  and  second hardest jet are shown for
the $pp\rightarrow t\bar{t}jj+ X$  process.  The blue dashed curve
corresponds to the leading order, whereas the red solid one to the
next-to-leading order result.  The histograms can also be turned into
dynamical K-factors, which we display in the lower panels.
Distributions demonstrate tiny corrections up to at least 200 GeV,
which means that the size of the corrections to the cross section is
transmitted to the distributions.  On the other hand, strongly altered
shapes are  visible at high $p_T$ especially in case of the first
hardest jet. Let us underline here, that corrections to the high $p_T$
region can only be correctly  described by higher order calculations
and are not altered by soft-collinear  emissions simulated by parton
showers.

\section{Summary}
We report the results of a next-to-leading order simulation of top quark pair 
production in association with two jets. With our inclusive cuts, we show 
that the corrections with respect to leading order are negative and small, 
reaching $11\%$. The error obtained by scale variation is of the same order.

\section{Acknowledgments}

I would like to thank the organizers of   the International Symposium
on Multiparticle Dynamics for the kind invitation and the very
pleasant atmosphere during the conference. 

The work presented here was
funded by the Initiative and Networking Fund of the Helmholtz
Association, contract HA-101 (“Physics at the Terascale”) and by the
RTN European Programme MRTN-CT-2006-035505 Heptools - Tools and
Precision Calculations for Physics Discoveries at Colliders.


\begin{footnotesize}

\end{footnotesize}


\end{document}